%Paper: hep-th/9508177
%From: Paul Townsend <P.K.Townsend@damtp.cam.ac.uk>
%Date: Thu, 31 Aug 95 16:57:21 BST
%Date (revised): Fri, 27 Oct 95 17:51:17 GMT
%Date (revised): Mon, 20 Nov 95 17:34:55 GMT

%%%%%%%This requires the PHYZZX.TEX macropackage

% Dyonic membranes
%J.M. Izquierdo, N.D. Lambert, G. Papadopoulos and P. K. Townsend

%%%%%%%If you do not have the msbm fonts, delete the following 4 lines
\font\mybb=msbm10 at 12pt
\def\bb#1{\hbox{\mybb#1}}
\def\Z {\bb{Z}}
\def\R {\bb{R}}
\def\E {\bb{E}}
%%%%%%%%%%%%
%%%and replace with the following 2 lines (without %)
%\def\Z {Z}
%\def\R {R}
%%%%%%%%%%

\tolerance=10000
\input phyzzx

 \def\unit{\hbox to 3.3pt{\hskip1.3pt \vrule height 7pt width .4pt \hskip.7pt
\vrule height 7.85pt width .4pt \kern-2.4pt
\hrulefill \kern-3pt
\raise 4pt\hbox{\char'40}}}

\def\cM {{\cal{M}}}

%%%%%%%%%%%%%%%%%%%%%%%%%%%%%%%%%%%%%%%%%%%%%%%%%%%%%%%%%%%%%%%%%%%%%%%%%%%%%
\REF\PKT{P.K. Townsend, {\it p-Brane Democracy}, hep-th/9507048, to appear in
proceedings of the PASCOS/Hopkins workshop, March 1995.}
\REF\Nep{R. Nepomechie, Phys. Rev. {\bf D31} (1984) 1921; C. Teitelboim, Phys.
Lett. {\bf B167} (1986) 69.}
\REF\DL{M.J. Duff and J.X. Lu, Nucl. Phys. {\bf B416} (1994) 301.}
\REF\DFKR{M.J. Duff, S. Ferrara, R.R. Khuri and J. Rahmfeld, Phys. Lett. {\bf
B356} (1995) 479.}
\REF\BBO{E. Bergshoeff, H.J. Boonstra and T. Ort{\'{\i}}n, {\sl S-Duality and
Dyonic $p$-brane Solutions in Type II String Theory}, hep-th/9508091.}
\REF\SaSe{A. Salam and E. Sezgin, Nucl. Phys. {\bf B258} (1985) 284.}
\REF\HT{C.M. Hull and P.K. Townsend, Nucl. Phys. {\bf B438} (1995) 109.}
\REF\Witb{E. Witten, Phys. Lett. {\bf B86} (1979) 283.}
\REF\GHT{G.W. Gibbons, G.T. Horowitz and P.K. Townsend, Class. Quantum Grav.
{\bf 12} (1995) 297.}
\REF\STW{A. Shapere, S. Trivedi and F. Wilczek, Mod. Phys. Lett. {\bf A6}
(1991)
2677.}
\REF\KO{R. Kallosh and T. Ort{\'{\i}}n, Phys. Rev. D {\bf 48} (1993) 742.}
\REF\PKTc{P.K. Townsend, Phys. Lett. {\bf 354B} (1995) 247.}
\REF\Witten{E. Witten, Nucl. Phys. {\bf B443} (1995) 85.}
\REF\FILQ{A. Font, L. Iba{\~n}ez, D. L{\"u}st and F. Quevedo, Phys. Lett. {\bf
B249} (1990) 35.}
\REF\Sen{A. Sen, Int. J. Mod. Phys. {\bf A 9} (1994) 3707; J.H. Schwarz, {\it
String Theory Symmetries}, hep-th/9503127.}
\REF\SS{J.H. Schwarz and A Sen, Phys. Lett. {\bf 312B} (1993) 105.}
\REF\Duff{M.J. Duff, Nucl. Phys. {\bf B442} (1995) 47.}
\REF\Ortin{T. Ort{\'{\i}}n, Phys. Rev. D {\bf 51} (1995) 790.}
\REF\PKTb{P.K. Townsend, Phys. Lett. {\bf 350B} (1995) 184.}
\REF\DS{M.J. Duff and K.S. Stelle, Phys. Lett. {\bf 253B} (1991) 113.}
\REF\Gu{R. G\"uven, Phys. Lett. {\bf 276B} (1992) 49.}
\REF\HTb{C.M. Hull and P.K. Townsend, Nucl. Phys. {\bf B451} (1995) 525.}

%%%%%%%%%%%%%%%%%%%%%%%%%%%%%%%%%%%%%%%%%%%%%%%%%%%%%%%%%%%%%%%%%%%%

\Pubnum{ \vbox{ \hbox{R/95/40}  \hbox{hep-th/9508177}} }
\pubtype{}
\date{August, 1995}

\titlepage

\title {\bf Dyonic membranes}

\author{J.M. Izquierdo, N.D. Lambert, G. Papadopoulos and P.K. Townsend}
\address{DAMTP, Silver St.,
\break
Cambridge CB3 9EW, U.K.}

\abstract{We present dyonic multi-membrane solutions of the N=2 D=8
supergravity
theory that serves as the effective field theory of the $T^2$-compactified type
II superstring theory. The `electric' charge is fractional for generic
asymptotic
values of an axion field, as for D=4 dyons. These membrane solutions are
supersymmetric, saturate a Bogomolnyi bound, fill out orbits of an $Sl(2;\Z)$
subgroup of the type II D=8 T-duality group, and are non-singular when
considered as solutions of $T^3$-compactified D=11 supergravity. On
$K_3$ compactification to D=4, the conjectured type II/heterotic equivalence
allows the $Sl(2;\Z)$ group to be reinterpreted as the S-duality group of the
toroidally compactified heterotic string and the dyonic membranes wrapped
around
homology two-cycles of $K_3$ as S-duals of perturbative heterotic string
states.}

\endpage
%\pagenumber=1

%%%%%%%%%%%%%%%%%%%%%%%%%%%%%%%%%CHAPTER 1%%%%%%%%%%%%%%%%%%%%%%%%%%%%%%%%%%%

\chapter{Introduction}

A feature of recent developments in superstring theory is the emerging
importance for a variety of non-perturbative phenomena of extended object, or
`$p$-brane', solutions of the classical string theory. In particular, these
solutions are crucial for an understanding of the various conjectured duality
symmetries of both the heterotic and type II superstrings (see [\PKT] for a
recent
review). It is customary to call a $p$-brane `electric' if it is the source for
a
$(p+1)$-form potential in the effective field theory Lagrangian and `magnetic'
if
it is the source for the dual $(D-p-3)$-form potential. The word `source' may
need
some explanation here: one first solves the source-free equations of motion of
the
effective field theory; it is necessary to introduce an actual, `fundamental',
source only if the analytic continuation of the source-free solution meets with
a
(timelike) singularity. Otherwise, no source is needed, but here one can
interpret
the extended object solution as an effective source on length scales that are
long
compared to the size of the object's core.

In a D-dimensional spacetime the magnetic dual of an electric $p$-brane is a
$\tilde p$-brane, where $\tilde p$ is related to $p$ by [\Nep]
$$
\tilde p = D-p-4\ .
\eqn\onea
$$
It follows that a $p$-brane can carry {\it both} electric and magnetic charge
only if
$$
D= 2p+4\ , \qquad p=0,1,2,\dots
\eqn\oneb
$$
The simplest case is $D=4$ for which there arises the possibility of particles
carrying both electric and magnetic charge, i.e. dyons. The next simplest case
is $D=6$ for which there exists the possibility of dyonic strings. In fact, one
can find a self-dual string in $D=6$, which is intrinsically dyonic because the
two-form potential to which it couples has a self-dual field strength [\DL].
Other dyonic D=6 strings, which break more than half the supersymmetry, have
been discussed in [\DFKR]. Here we consider the next case: membranes in D=8.
Specifically, we present dyonic membrane solutions of N=2 D=8 supergravity that
break half the supersymmetry.  During the writing up of this work a paper
[\BBO] presenting analogous results for D=6 dyonic strings appeared, in which
the possibility of D=8 dyonic membranes was also mentioned.
The reason for considering N=2 D=8 supergravity [\SaSe] is that this is the
unique supersymmetric field theory (with no more than second order field
equations) for which the field content includes a third-rank antisymmetric
tensor gauge field. It may also be considered as the effective field theory for
the $T^2$-compactified type II superstring.

The N=2 D=8 supergravity theory has an $Sl(3;\R)\times Sl(2;\R)$ symmetry of
the equations of motion. This group acts linearly on the various field strength
tensors and their duals. These include a four-form field-strength $F$, and its
dual,
which transform according to the $({\bf 1},{\bf 2})$ representation of
$Sl(3;\R)\times
Sl(2;\R)$. The discrete subgroup $Sl(3;\Z)\times Sl(2;\Z)$ was conjectured in
[\HT] to
extend to a U-duality of the D=8 type II superstring theory; this discrete
group
contains the T-duality group $SO(2,2;\Z)\equiv [Sl(2;\Z)\times Sl(2;\Z)]/Z_2$,
which
in turn contains an $Sl(2;\Z)$ subgroup of the $Sl(2;\R)$ group acting on the
four-form field strength. This follows from the facts that (i) all
non-perturbative
U-duality symmetries are contained in the $Sl(3;\Z)$ subgroup
and (ii) $Sl(3;\Z)$ acts trivially on $F$. Thus, although the $Sl(2;\Z)$ group
acts on $F$ via a generalized electromagnetic duality, this group is
nevertheless a
{\it perturbative} T-duality in the string theory context. This is to be
expected
from the fact that in the context of the type II superstring the three-form
potential $A$ is a Ramond-Ramond (RR) field. A similar group-theoretical
argument was
used in [\HT] to show that electric {\it and} magnetic RR charges of the D=4
type II
superstring transform irreducibly under T-duality. As we now see, the same is
true in D=8.

It should be noted that here we are using the term `T-Duality', in the context
of $T^2$-compactifications of type II superstrings, to mean the identification
of vacua of the resulting D=8 type II superstring under the discrete
$SO(2,2;\Z)$ subgroup of the $SO(2,2)$ classical symmetry group of
the compactified theory. The analogous $SO(2,18;\Z)$ T-Duality group of the
heterotic string includes transformations which take $R\rightarrow 1/R$, where
$R$ is the radius of an $S^1$ factor of $T^2$. For the type II superstrings
this $R\rightarrow 1/R$ transformation (also called T-Duality) interchanges the
type IIA and type IIB superstrings, which are therefore equivalent to a single
D=8 type II superstring. Such transformations are {\it not} realized as gauge
symmetries of this D=8 theory, and therefore are not included in the
$SO(2,2;\Z)$ T-Duality group.

There is a consistent truncation of the N=2 D=8 supergravity in
which the only surviving fields are the spacetime metric, $g_{\mu\nu}$, a
scalar, $\sigma$, a pseudoscalar $\rho$ and a three-form gauge potential, $A$,
for which $F=dA$ is its four-form field-strength. The Lagrangian of this
truncated
theory is
$$
\eqalign{
{\cal L} = N\Bigg\{\sqrt{-g}\big[ &R - 2\partial_\mu \sigma\partial^\mu\sigma -
2e^{4\sigma}\partial_\mu \rho\partial^\mu\rho  -{1\over
12}e^{-2\sigma}F_{\alpha\beta\gamma\delta}F^{\alpha\beta\gamma\delta}\big] \cr
&-{1\over 144}\varepsilon^{\mu\nu\rho\sigma\alpha\beta\gamma\delta}\rho
F_{\mu\nu\rho\sigma}F_{\alpha\beta\gamma\delta} \Bigg\}\ ,}
\eqn\onec
$$
where $N$ is a normalization factor, which we can choose at our convenience.
The coefficient of the $\varepsilon\rho FF$ term is crucial to the results to
follow so we should point out that we disagree by a factor of three with the
coefficient of this term given in [\SaSe]. The coefficient can be simply
determined by dimensional reduction of the D=11 supergravity theory, which was
the method used in [\SaSe], but this leads to the coefficient used here rather
than that of [\SaSe].

The $\sigma$ and $\rho$ kinetic terms of \onec\ constitute a
sigma model with target space $Sl(2;\R)/U(1)$. It is convenient to introduce
the
complex field
$$
\lambda = 2\rho + ie^{-2\sigma}\ ,
\eqn\comfield
$$
taking values in the upper half complex plane, since the $Sl(2;\R)$ group acts
on
$\lambda$ by fractional linear transformations. Since the asymptotic value of
$\lambda$ is undetermined by the equations of motion, the possible vacua
correspond to points in the upper half plane. However, T-duality of the
type II D=8 superstring theory implies that points that lie in an
orbit of an $Sl(2;\Z)$ subgroup of $Sl(2;\R)$ correspond to equivalent vacua.
Thus, the moduli space of vacua in the string theory context is, assuming
T-duality, the fundamental domain of $SL(2;\Z)$ in the upper half complex
plane.

Note the similarity of the above discussion with that of S-duality in the
heterotic string. The main difference, apart from the obvious one that here we
are
dealing with a four-form rather than a two-form field strength, is that the
scalar
field $\sigma$ in \onec\ is {\it not} the dilaton. In fact, the dilaton has
been set to zero in the truncation leading to \onec. If $\sigma$ were the
dilaton, the $Z_2$ subgroup of $Sl(2;\Z)$ that exchanges $F$ with its dual and
takes $\sigma$ to $-\sigma$ would be
non-perturbative in the context of the type II string theory. As explained
above, this is not the case. An alternative explanation is provided by
string-string duality, as will shall see shortly.

We shall be interested in infinite planar membrane solutions of the equations
of
motion of \onec\ that are asymptotically flat as one approaches spatial
infinity
in non-coplanar directions; we shall call this `transverse spatial infinity',
which is topologically $S^4\times \R^2$. Membrane solutions can be
characterised
by their electric and magnetic number densities
$$
q = {N\over e}\oint\! G \qquad p= {e\over 2\pi}\oint\! F
\eqn\chargetwo
$$
where the integral is over a 4-sphere cross-section of transverse spatial
infinity, $e$ is an arbitrary unit of `electric' charge, and the two-form $G$
is
related to the Hodge dual $\tilde F$ of $F$ by
$$
G\equiv e^{-2\sigma}\tilde F -2\rho F \ .
\eqn\chargetwo
$$
We shall require an asymptotic translational invariance in directions coplanar
with the membrane so that these number densities are actually constant; we
shall
refer to these constants as the membrane `charges'. Their conservation follows
from the fact that the combined equations of motion and Bianchi identities of
the
field-strength four-form $F$ can be written as $d{\cal F}=0$ where ${\cal F}$
is
the $Sl(2;\R)$ doublet
$$
{\cal F} = (F,G)\ .
\eqn\chargethree
$$
We shall choose the constants $N$ and $e$ such that
$$
q = {1\over \Omega_4}\oint G \qquad p= {1\over \Omega_4}\oint F
\eqn\charge
$$
where $\Omega_4=2\pi^2$ is the volume of the unit 4-sphere. With this choice,
the
charges $(p,q)$ form an $Sl(2;\R)$ doublet.

As shown in [\Nep], the electric and magnetic charges of extended objects are
subject to a generalization of the Dirac quantization condition. However, just
as
the Dirac quantization condition must be replaced, in the context of dyons, by
the Schwinger-Zwanziger quantization condition so, in the context of dyonic
extended objects, the Nepomechie-Teitelboim (N-T) quantization condition must
be
replaced by an extended object analogue of the Schwinger-Zwanziger quantization
condition. With the above choice of normalization constant, $N$, and electric
charge unit, $e$, this generalized N-T quantization condition for two dyonic
membranes with charges $(p,q)$ and $(p',q')$ takes the simple (manifestly
$Sl(2;\R)$ invariant) form
$$
qp'-q'p \ \in \ \Z\ .
\eqn\Dizzy
$$
As for dyons in D=4 [\Witb], this formula allows fractional $q$ for dyonic
membranes, but the consequences for dyonic membranes are not quite the same as
those for dyons because one cannot take for granted the existence of purely
electric membranes in the quantum theory.

In [\GHT] it was shown how an analogue of the Bogomolnyi-Gibbons-Hull bound for
particle-like solutions of Maxwell/Einstein theory can be derived for $p$-brane
solutions of certain antisymmetric tensor generalizations of Maxwell/Einstein
theory. The precise interactions of the antisymmetric tensor field, e.g. the
coefficient of possible Chern-Simons terms was crucial to this result. In all
cases,
the interactions were precisely those for which the bosonic field theory could
be
interpreted as a consistent truncation of a supergravity theory. Since this
condition is satisfied by the Lagrangian \onec\ one would expect to be able to
derive a similar bound on the tension of membrane solutions of its equations of
motion; this case is not covered by the results of [\GHT] because Lagrangians
with
scalar fields were not considered there. This expectation is correct; we shall
show
that the tension, $M$, of membrane solutions of \onec\ satisfies the $Sl(2;\R)$
invariant bound
$$
M^2 \ge {1\over 4} \Big[ e^{2\langle\sigma\rangle}\big(q+2\langle\rho\rangle
p\big)^2 + e^{-2\langle\sigma\rangle} p^2\Big] \ ,
\eqn\abog
$$
where $\langle \rho\rangle$ and $\langle\sigma\rangle$ are the asymptotic
values
of $\rho$ and $\sigma$.

Solutions which saturate the bound are `supersymmetric' in that they admit
Killing
spinors. The purely electric and magnetic D=8 supersymmetric membrane
solutions, with
$\rho\equiv0$, have been given previously [\DL]. The supersymmetric membrane
solutions we construct here differ in that they have non-constant axion field
and
carry both electric and magnetic charge, i.e. they are `dyonic'. There is a
U(1)
parameter family of these solutions for each value of the asymptotic values of
$\sigma$ and $\rho$, corresponding to the U(1) stability subgroup of
$Sl(2;\R)$ acting on the upper-half plane by fractional linear transformations.
Although only a $Z_2$ family of these will survive quantization, the
identification
of vacua related by a transformation in the $Sl(2;\Z)$ T-duality subgroup of
$Sl(2;\R)$ allows us to find $Sl(2;\Z)$ orbits of membrane solutions about
equivalent vacua, as has been done previously for particle-like solutions in
D=4
[\STW,\KO].  Almost all such solutions are dyonic.

One motivation for our work derives from a recently suggested D=8
membrane/membrane
duality [\PKTc]. The point here is, firstly, that while the purely electric
membrane
solution of N=2 D=8 supergravity theory can be interpreted as the membrane
solution
of D=11 supergravity in a $T^3$ compactified spacetime, the purely magnetic one
can
be interpreted as a double dimension reduction of the fivebrane solution of
D=11
supergravity\foot{This was stated in [\PKTc]; here we verify it.}. Secondly,
the
worldvolume action of this magnetic membrane is that of a D=11 supermembrane in
a
$T^3$ compactified spacetime (and not that of a D=8 supermembrane, as one might
have guessed; the extra three coordinates come from the antisymmetric tensor in
the fivebrane's worldvolume action). This suggests a complete non-perturbative
equivalence between the electric and magnetic membranes. This equivalence would
be
guaranteed in string theory by non-perturbative T-duality. Unfortunately, this
cannot be established in string perturbation theory, but one can reverse the
logic
and use the evidence of membrane/membrane duality given in [\PKTc] and the
results
presented here as evidence for the non-perturbative validity of T-duality.

Another motivation comes from the conjectured non-perturbative
equivalence of the $K_3\times T^2$ compactified type II superstring theory with
the toroidally compactified heterotic string theory [\HT], i.e. the
`string-string
duality' for which there is now considerable evidence. Many recent papers
dedicated to
tests of this conjecture  have taken as their starting point the related
conjecture
that the D=6 string theories obtained by compactification of the type IIA
superstring
on $K_3$ and the heterotic string on $T^4$ are non-perturbatively equivalent
[\Witten]. Given this D=6 equivalence, the equivalence in D=4 follows upon
further
compactification on $T^2$. S-duality of the heterotic string [\FILQ,\Sen] can
then be
re-interpreted as T-duality of the type II superstring [\SS,\Duff,\Witten].
This
approach to understanding D=4 S-duality via the heterotic/type II equivalence
can be
characterised  by the motto ``10 to 6 and then to 4''.

Our work can be viewed as
a first step towards an understanding of heterotic S-duality via the
alternative
``10 to 8 and then to 4'' approach. The first step is a $T^2$ compactification
of
both the type II and the heterotic string to D=8. A subsequent compactification
of
the D=8 type II superstring on $K_3$ yields a D=4 string theory which,
according to
string-string duality, is equivalent to the $T^4$ compactified D=8 heterotic
string.
The spectrum of this D=4 string theory includes dyons which arise, in the type
II
interpretation, as wrapping modes of D=8 dyonic membranes around the 22
fundamental
homology 2-cycles of $K_3$. These dyons are charged with respect to the 22 D=4
two-form field strengths, $F^I\  (I=1,2,\dots,22)$, arising from the D=8
four-form
field strength $F$ via the ansatz
$$
F(x,y)= F^I(x)\wedge\omega_I(y) \ ,
\eqn\ansatz
$$
where $\omega_I$ span the 22-dimensional space of harmonic two-forms on $K_3$.
These
D=4 dyons are non-perturbative RR states, even the purely electric ones; they
form multiplets of the $Sl(2;\Z)$ (type II) T-duality subgroup descending from
the
T-duality group in D=8. Note that the full type II T-duality group in D=4 is
the same
as the full type II T-duality group in D=8 because $K_3$ has no continuous
isometries.

According to string-string duality the type II RR dyons just discussed must
appear
in the spectrum of an equivalent heterotic string. Moreover, one expects the
purely electric
particles among them to appear as {\it perturbative} states in view of the
generally accepted opinion that {\it all} purely electric states of the
heterotic
string are perturbative. Since their dyonic $Sl(2;\Z)$ partners are necessarily
non-perturbative, the $Sl(2;\Z)$ group that relates them must then be a
{\it non-perturbative} duality group of the heterotic string, i.e. the
S-duality group. Thus, the existence of the $Sl(2;\Z)$ multiplets of dyonic D=8
membranes provides further confirmation of the interchange of S and T duality
effected by string-string duality.

In the following, we begin with a presentation of the dyonic membrane solutions
of
the field equations of the Lagrangian \onec. We then explain how these
solutions were found and why their tension saturates a Bogomolnyi-Gibbons-Hull
type
bound. We also exhibit the Killing spinors admitted by these solutions,
thereby establishing their supersymmetry. We then discuss the global
structure of the dyonic membranes and their interpretation as solutions of D=11
supergravity. We conclude with some further comments on the significance of our
results.

%%%%%%%%%%%%%%%%%%%%%%%%%%%%%%%%%%Chapter 2%%%%%%%%%%%%%%%%%%%%%%%%%%%%%%%%%%

\chapter{D=8 dyonic membranes}

The field equations of the Lagrangian \onec\ are
$$
\eqalign{
G_{\mu\nu} &= 2T_{\mu\nu}
\cr
\partial_\mu \big(\sqrt{-g}\; e^{-2\sigma}F^{\mu\nu\rho\sigma}\big) &=-
2\big(\partial_\mu\rho\big)
\tilde F^{\mu\nu\rho\sigma}
\cr
\partial_\mu\big(\sqrt{-g}\; e^{4\sigma}\partial^\mu\rho \big) &= {1\over 24}
F_{\mu\nu\rho\sigma}\tilde F^{\mu\nu\rho\sigma}
\cr
\partial_\mu\big(\sqrt{-g}\; \partial^\mu\sigma\big) &= \sqrt{-g}\big[2
e^{4\sigma}(\partial\rho)^2 -{1\over24} e^{-2\sigma} F^2 \big]\ ,}
\eqn\aonec
$$
where
$$
\eqalign{
T_{\mu\nu} = &\big[\partial_\mu\sigma\partial_\nu\sigma - {1\over2}
g_{\mu\nu}(\partial\sigma)^2\big] + e^{4\sigma}
\big[\partial_\mu\rho\partial_\nu\rho - {1\over2}
g_{\mu\nu}(\partial\rho)^2\big]\cr
&\qquad +{1\over6}e^{-2\sigma} \big[
F_{\mu\alpha\beta\gamma}F_{\nu}{}^{\alpha\beta\gamma} -{1\over8}g_{\mu\nu}F^2
\big] \ ,}
\eqn\bonec
$$
and
$$
\tilde F^{\mu\nu\rho\sigma} \equiv {1\over 24}
\varepsilon^{\mu\nu\rho\sigma\alpha\beta\gamma\delta}
F_{\alpha\beta\gamma\delta}\ .
\eqn\donec
$$
We shall consider field configurations representing an infinite planar
membrane and choose coordinates such that it is aligned with the $x^1\equiv y$
and $x^2\equiv z$ axes. We shall look for product metrics in which the metric
of the
five-dimensional `transverse' space is conformally flat and may therefore be
parameterised by the coordinates ${\bf x}\equiv (x^3,\dots,x^7)$ of an
associated
five-dimensional Euclidean space, $\E^5$. There are certainly many solutions of
the
field equations \onec\ within this class of field configurations, but we shall
concentrate on those that admit Killing spinors. We shall first present these
solutions. Then, in the following section, we shall explain how they were
obtained
and why they are supersymmetric. We shall present the solutions in terms of
the complex field $\lambda$ defined in \comfield. If we fix boundary conditions
such that the spacetime is asymptotically flat as $|{\bf x}|\rightarrow
\infty$, and such that
$$
\lambda \rightarrow i\ ,
\eqn\bcs
$$
then the following multi-membrane field configurations solve \aonec\ for
arbitrary angular parameter $\xi$:
$$
\eqalign{
ds^2 &= H^{-{1\over2}}[-dt^2 + dy^2 + dz^2] + H^{1\over2} d{\bf
x}\cdot d{\bf x}\cr
F &= {1\over2}\cos\xi\, (\star dH) + {1\over2}\sin\xi\, dH^{-1}\wedge dt\wedge
dy\wedge dz
\cr
\lambda &= {\sin 2\xi\,(1-H) + 2iH^{1\over2}\over 2 (\sin^2\xi + H\cos^2\xi)}\
.}
\eqn\dyons
$$
Here, the symbol $\star$ indicates the Hodge dual in $\E^5$ and
$$
H= 1 + \sum_{n=1}^N{\mu_n\over |{\bf x}-{\bf x}_n|^3}
\eqn\harm
$$
for $n$ arbitrary constants $\mu_n$ associated with the $N$ points  ${\bf
x}={\bf x}_n$, for any finite value of $N$. That is, $H({\bf x})$ solves the
Laplace equation on $\E^5$ with an arbitrary number of point sources
and is such that
$H\rightarrow 1$ as $|{\bf x}|\rightarrow \infty$. The constants $\mu_n$ are
proportional to the ADM tension of each membrane solution. Specifically, for a
one membrane solution with parameter $\mu$ the ADM tension is
$$
M= {3\over4}\mu\ .
\eqn\tension
$$

We have presented the solutions for a specially chosen asymptotic value of
$\lambda$ because a solution with any other asymptotic value of $\lambda$ can
be
found by making use of the $Sl(2;\R)$ invariance of the field equations. As
stated earlier, this $Sl(2;\R)$ group acts on $\lambda$ by fractional linear
transformations:
$$
\lambda \rightarrow {a\lambda + b\over c\lambda + d}\ ,
\eqn\onee
$$
where $a,b,c,d$ are real numbers such that $ad-bc=1$. The $Sl(2;\R)$ group acts
on the four-form doublet ${\cal F} = (F, G)$ by a generalization of
electromagnetic duality. Specifically, if $\lambda$ is transformed as in \onee,
then the associated transformation of ${\cal F}$ is
$$
{\cal F}\rightarrow (F, G) \pmatrix{d&-b\cr -c& a}\ .
\eqn\onefb
$$

Since there is a $U(1)$ isotropy subgroup of $Sl(2;\R)$ that does not change
the
asymptotic value, $\langle\lambda\rangle$, of $\lambda$, there must be a $U(1)$
family of solutions for each choice of $\langle\lambda\rangle$. This is the
significance of the angular parameter $\xi$ in \dyons. This $U(1)$ group
is an analogue of the electromagnetic duality group since it takes a purely
electric or purely magnetic solution into a dyonic one. Thus, the general
solution of \dyons\ can be obtained by a $U(1)$ transformation of the purely
magnetic solution
$$
\eqalign{
ds^2 &= H^{-{1\over2}}[-dt^2 + dy^2 + dz^2] + H^{1\over2} d{\bf
x}\cdot d{\bf x}\cr
F &= {1\over2}\star dH \cr
\lambda &= iH^{-{1\over2}} \ .}
\eqn\dyonstwo
$$

However, because of charge quantization, this classical $U(1)$ symmetry will be
broken to $Z_2$ in the quantum theory; there will be some `preferred' value of
$\langle\lambda\rangle$ for which only the purely electric or purely magnetic
solutions survive (by analogy with D=4 dyons one might suppose that
$\langle\lambda\rangle =i$ is the `preferred' value; we shall examine this
hypothesis in more detail later). It might therefore appear that the more
general
dyonic membrane solutions of \dyons\ are irrelevant to the type II string
theory,
at least for the `preferred' value of $\langle\lambda\rangle$. However, the
sigma-model target space of \onec\ is only required by supersymmetry to be {\it
locally} isometric to the coset space $SL(2;\R)/U(1)$. It may differ globally
since
it is possible to identify points on this space that differ by the action of
$Sl(2;\Z)$. Thus, the true sigma-model space could be
$$
\cM = Sl(2;\Z)\backslash Sl(2;\R)/U(1)\ .
\eqn\true
$$
In this case the true moduli space is not the entire upper-half $\lambda$-plane
but
rather the fundamental domain of $Sl(2;\Z)$ in the upper half plane. In the
context
of the D=8 type II superstring theory, T-duality implies that this is indeed
the
true moduli space of vacua, so vacua which  differ by the action of $Sl(2;\Z)$
should be identified. Thus an $Sl(2;\Z)$ transformation of
the purely magnetic membrane solution \dyonstwo\ will produce a new solution
with a
different, but {\it equivalent}, value of $\lambda$, and this solution will
have
an effective non-zero value of $\xi$, i.e. it will be dyonic.

Actually, we shall find a more general class of dyonic solutions by applying
this procedure to the dyonic solutions \dyons\ rather than to the purely
magnetic
solution \dyonstwo, i.e. we allow for an arbitrary initial value of the angular
parameter $\xi$. First we make an $Sl(2;\R)$ transform of the solution \dyons\
to arrive at
$$
\eqalign{
ds^2 &= H^{-{1\over2}}[-dt^2 + dy^2 + dz^2] + H^{1\over2} d{\bf
x}\cdot d{\bf x}\cr
F &= {1\over2}e^{2\langle\sigma\rangle}\Big( \cos\psi \star dH + \sin\psi\;
dH^{-1} \wedge dt\wedge dy\wedge dz\Big)
\cr
\lambda &= 2\langle \rho\rangle + e^{-2\langle\sigma\rangle} \cdot {(1-H)\sin
2\psi + 2i H^{1\over2}\over 2(H\cos^2\psi + \sin^2\psi)}
\ ,}
\eqn\dyonsthree
$$
where
$$
e^{-2\langle\sigma\rangle} = {1\over c^2 + d^2}\ ,\qquad
2\langle\rho\rangle = {bd + ac\over c^2 + d^2}\ ,
\eqn\dyonsfour
$$
and the new angular parameter $\psi$ is given by
$$
\tan\psi = {d\sin\xi + c\cos\xi\over d\cos\xi -c\sin\xi}\ .
\eqn\adyonsfour
$$
Then, we restrict $a,b,c,d$ to be integers to obtain the dyon solutions with
$\langle\lambda\rangle \cong i$. By construction, these solutions form a
representation of $Sl(2;\Z)$. Note that the set of dyon solutions obtained in
this
way will contain a purely magnetic solution if and only if $\tan\xi$ is
rational.
If this condition is satisfied then there will also be a purely electric
solution.

Clearly, a similar set of dyonic membrane solutions can be found for any other
initial choice of $\langle\lambda\rangle$. However, if initially
$\langle\lambda\rangle \ne i$, then the $Sl(2;\Z)$ subgroup is not found by
simply
restricting $a,b,c,d$ to be integers. Rather, the elements of the $Sl(2;\Z)$
subgroup are similarity transforms of matrices with integer entries.

%%%%%%%%%%%%%%%%%%%%%%%%%%%%%%%%Chapter 3%%%%%%%%%%%%%%%%%%%%%%%%%%%%%%%%%%%

\chapter{Killing spinors and the Bogomol'nyi Bound}

We have claimed that the dyonic membrane solutions presented above are
supersymmetric, i.e. that they admit Killing spinors. We shall now elaborate on
this point. A Killing spinor is a spinor field, $\epsilon$, that is in the
kernel
of a first-order Lorentz-covariant Dirac-type operator $\hat{\cal D}$, i.e.
$\hat{\cal D}\epsilon=0$, where a minimal condition on $\hat{\cal D}$ is that
the
vector field  $\bar\epsilon\gamma^\mu\epsilon$ is  Killing if $\epsilon$ is. In
the
context of field theories with scalar and vector fields,  this condition
limits,
but does not define, $\hat{\cal D}$. Within the context of a supergravity
theory,
$\hat{\cal D}$ is defined by the gravitini transformation laws, but an
alternative
intrinsic definition is possible in the context of an {\sl a priori} arbitrary
bosonic Lagrangian via the modified  Nester tensor
$$
\hat E^{\mu\nu} = {1\over2}\bar\epsilon\Gamma^{\mu\nu\rho}\hat{\cal
D}_\rho\epsilon + c.c.\ .
\eqn\Nester
$$
This is because the operator $\hat{\cal D}$ is fixed, if it exists, by the
requirement that
$$
{\cal D}_\nu\hat E^{\mu\nu} = \overline{\hat{\cal D}_\nu\epsilon}\;
\Gamma^{\mu\nu\rho}\hat{\cal D}_\rho\epsilon - {1\over2} \bar\chi
\Gamma^\mu\chi\ ,
\eqn\crucial
$$
as a consequence of the field equations, for some complex spinor $\chi$. This
requirement also fixes $\chi$. The significance of the relation \crucial\ is
that it allows the derivation of a bound on the mass per unit $p$-volume,
i.e. the tension, of configurations that are subject only to the boundary
conditions at transverse spatial infinity satisfied by $p$-brane
solutions of the equations of motion [\GHT]. It can happen that the field
equations of a
given Lagrangian are such that \crucial\ is not satisfied by any operator
$\hat{\cal D}$
for any spinor $\chi$. In this case a bound on the tension cannot be derived by
this method. Conversely, requiring that such a bound be derivable in a
Lagrangian whose interactions are parameterised by arbitrary functions of the
scalar fields can fix these functions. For example, allowing arbitrary
interactions
of $\sigma$ consistent with the requirement that the field equations be of
second
order, and an arbitrary coefficient of the $\rho F\tilde F$ term, one finds
that the only Lagrangian in this class for which an energy bound on the
membrane tension can be derived is precisely the Lagrangian of \onec.

For the case in hand, one finds that
$$
\hat{\cal D}_\mu \epsilon \equiv {\cal D}_\mu\epsilon -{1\over2}
\gamma_9\epsilon\; e^{2\sigma}
\partial_\mu \rho + {1\over 96}\Gamma^{\alpha\beta\gamma\delta}
\Gamma_\mu\epsilon\;  e^{-\sigma}
F_{\alpha\beta\gamma\delta}\ ,
\eqn\oneh
$$
and
$$
\chi = \Gamma^\mu\epsilon\; \partial_\mu\sigma -
\gamma_9\Gamma^\mu\epsilon\; e^{2\sigma}\partial_\mu\rho -{1\over 48}
\Gamma^{\alpha\beta\gamma\delta}\epsilon\;
e^{-\sigma}F_{\alpha\beta\gamma\delta}\ .
\eqn\onei
$$
The matrix $\gamma_9$ is defined by
$$
\gamma_9 =\Gamma^{\underline 0}\Gamma^{\underline 1}\cdots\Gamma^{\underline 7}
\eqn\onej
$$
where the underlining indicates a flat space Dirac matrix. It follows from
\oneh\
that
$$
\hat E^{\mu\nu} = E^{\mu\nu} -{1\over2} e^{2\sigma} (\bar\epsilon
\Gamma^{\mu\nu\alpha}\gamma_9 \epsilon)\partial_\alpha \rho - {1\over4}
e^{-\sigma}\bar\epsilon(F^{\mu\nu\alpha\beta}\Gamma_{\alpha\beta} -\tilde
F{}^{\mu\nu\alpha\beta}\Gamma_{\alpha\beta}\gamma_9)\epsilon
\eqn\aonej
$$
where $E^{\mu\nu}$ is the standard Nester tensor. Note that the Dirac
conjugate $\bar\psi$ of a spinor $\psi$ is defined by
$$
\bar\psi = \psi^{\dagger} \Gamma^{\underline 0}\ ,
\eqn\conjugate
$$
so that $\bar\psi\Gamma^{\underline 0}\psi$ is negative definite. Note also
that
the Lorentz invariant $\bar\psi\psi$ is pure imaginary (for commuting spinors)
since $\Gamma^{\underline 0}$ is anti-Hermitian\foot{In the Majorana basis, in
which the matrices $\Gamma^\mu$ are (for D=8) pure imaginary,
$\Gamma^{\underline 0}$ is symmetric and equal to $i$ times the charge
conjugation matrix (which is symmetric for D=8).}

As explained in the introduction, the relevant concept for defining
membrane charges is transverse spatial infinity, which has topology $S^4\times
\R^2$. It is convenient to choose periodic boundary conditions to convert this
to
$S^4\times T^2$, i.e. we consider the membrane to be wrapped around a large
two-torus. The energy per unit area, $M$, is then the, now finite, total energy
divided by the volume, $V_2$, of the two-torus. This energy can expressed as an
integral over the $S^4\times T^2$ surface at spatial infinity. Specifically, if
${\bf P}$ is the total transverse 5-momentum per unit area, such that
$M=\sqrt{-|{\bf P}|^2}$, then [\GHT]
$$
\bar\epsilon_\infty {\bf \Gamma}\cdot {\bf P}\epsilon_\infty = {1\over 2V_2
\Omega_4}
\oint_\infty \! dS_{\mu\nu} E^{\mu\nu}\ ,
\eqn\boga
$$
where $\Omega_4$ is the volume of the unit 4-sphere. With appropriate
asymptotic
fall off conditions on the metric, and assuming that
$$
\epsilon  \rightarrow \epsilon_\infty
\eqn\eplim
$$
as $|{\bf x}|\rightarrow \infty$, for some constant spinor $\epsilon_\infty$,
\boga\ can be rewritten as
$$
\bar\epsilon_\infty {\bf \Gamma}\cdot {\bf P}\epsilon_\infty = {1\over 2
\Omega_4}
\oint_\infty \! dS_{ij} E^{ij}\ ,
\eqn\bogb
$$
where the integral is now over the 4-sphere at spatial infinity and the index
$i$
is associated with the coordinates ${\bf x}$ of the transverse space.

Assuming that the only components of $F$ that are non-vanishing at transverse
spatial infinity are $F_{ijkl}$ and $F_{tyzi}$, and that these components
depend
asymptotically only on $x^i$, one has that
$$
\eqalign{
{1\over 2V_2\Omega_4} \oint_\infty \! dS_{\mu\nu}\hat E^{\mu\nu} &=
{1\over 2 \Omega_4} \oint_\infty \! dS_{ij} \hat E^{ij}\cr
&= \bar\epsilon_\infty \Big[{\bf \Gamma}\cdot {\bf P} -
{1\over 8 \Omega_4} e^{-\langle\sigma\rangle} \Gamma_{kl}\oint_\infty \!
dS_{ij}\Big( F^{ijkl} - \tilde F^{ijkl}\gamma_9\Big)\Big]\epsilon_\infty\ , }
\eqn\abog
$$
since the $\partial\rho$ term in \aonej\ does not contribute to the integral.
{}From
the definitions \charge\ of the charges $(p,q)$ one then finds that
$$
{1\over 2V_2\Omega_4} \oint_\infty \! dS_{\mu\nu}\hat E^{\mu\nu} =
\bar\epsilon_\infty K \epsilon_\infty
\eqn\bogc
$$
where
$$
K= {\bf \Gamma}\cdot {\bf P} -{1\over2} \Big[
e^{\langle\sigma\rangle}(q+2\langle\rho\rangle p)
\Gamma_{yz} - e^{-\langle\sigma\rangle} p\;
\Gamma_{yz}\gamma_9\Big]\ .
\eqn\abogc
$$
Using Gauss's law, the relation \crucial, and choosing $\epsilon$ to satisfy a
`modified Witten condition', one can prove that the integral on the left hand
side
of \bogc\ is positive semi-definite, subject to the usual assumptions. It
follows that the Dirac
matrix $K$ is positive semi-definite, which implies the bound \abog\ quoted in
the
introduction.

This bound is saturated by solutions of the equations of motion for which there
exists a spinor $\epsilon$ such that
$$
\hat{\cal D}_\mu\epsilon=0\ , \qquad\qquad \chi=0\ .
\eqn\Killing
$$
Non-trivial solutions of these relations, i.e. those for which $M\ne0$, require
$\epsilon$ to satisfy a condition of the form
$$
\big[\alpha({\bf x}) \Gamma_* + \beta({\bf
x})\Gamma_*\gamma_9\big]\epsilon({\bf x}) =\epsilon ({\bf x})\ ,
\eqn\constraint
$$
where
$$
\Gamma_* = \Gamma^{\underline 0}\Gamma^{\underline 1}\Gamma^{\underline 2}\ .
\eqn\onek
$$
and $\alpha$ and $\beta$ are functions such that
$$
\alpha^2 + \beta^2=1 \ .
\eqn\aonek
$$
This can be seen from the fact that the spinor $\epsilon$ must be an
eigen-spinor
of the matrix $K$ with zero eigenvalue. The angular parameter $\xi$ enters into
the
solutions \dyons\ as the limit of the ratio of the functions $\alpha$ and
$\beta$,
i.e.
$$
\lim_{|{\bf x}|\rightarrow \infty} \Big({\alpha\over\beta}\Big) = \tan\xi\ .
\eqn\bonek
$$

The multi-dyon solutions \dyons\ were obtained by substituting an appropriate
ansatz into the relations \Killing. The constraint \constraint\ reduces the
dimension of the space of Killing spinors to half that of the constant Killing
spinors of the vacuum. Thus, the solutions we find in this way will break half
the
supersymmetry. Furthermore, they saturate the bound \abog\ by construction, so
their membrane tension is given by the formula
$$
M^2 = {1\over 4} \Big[ e^{2\langle\sigma\rangle}(q+2\langle\rho\rangle p)^2 +
e^{-2\langle\sigma\rangle} p^2\Big] \ .
\eqn\bogi
$$
where $M$ is related to the constants $\mu_n$ appearing in the solutions by
$$
M= {3\over4} \sum_{n=1}^N \mu_n\ .
\eqn\bogsum
$$
This bound does not imply that each constant $\mu$ is individually positive but
it
is easy to see that this must in fact be the case. The point is that \bogsum\
is
independent of the positions of the membranes, so that we may consider a limit
in
which they become arbitrarily far apart. In this limit we may also take the
infinite radius limit of a four-sphere surrounding any given membrane of
tension
$\mu$. This four-sphere approaches transverse spatial infinity in this limit
but
encloses only the chosen membrane, so that $\mu$ must be positive. Because of
this,
the only singularities of the metric are at the `centres' ${\bf x}={\bf x}_n$.
The question whether these are real singularities or merely coordinate
singularities will be addressed in the following section. From
\onefb\ we  see that the $Sl(2;\R)$ transformation of (p,q) is
$$
(p,q)\rightarrow (p,q)\pmatrix{d&-b\cr -c& a}\ .
\eqn\abogsum
$$
Given that $\langle\sigma\rangle$ and $\langle\rho\rangle$ are also transformed
according to \onee, the $SL(2;\R)$ invariance of the formula \bogi\ is easily
verified.

The above procedure has the advantage that it not only
yields the solutions admitting Killing spinors, for given boundary conditions,
but
also the Killing spinors. For the solutions \dyons\ one finds that
$$
\eqalign{
\epsilon &= {1\over\sqrt{2}} H^{-{1\over8}}(H\cos^2\xi +
\sin^2\xi)^{-{1\over4}}\Big\{
\big[ (\sin^2\xi + H\cos^2\xi)^{1\over2} + H^{1\over2}\cos\xi \big]^{1\over2} +
\cr
&\qquad \big[ (\sin^2\xi + H\cos^2\xi)^{1\over2} - H^{1\over2}\cos\xi
\big]^{1\over2}\gamma_9\Big\}\epsilon_0\ ,}
\eqn\onej
$$
where the constant spinor $\epsilon_0$ must satisfy
$$
\Gamma_*\gamma_9\epsilon_0 = \epsilon_0\ ,
\eqn\aonej
$$
in order that $\epsilon$ satisfy the constraint \constraint.
It follows that the dimension of the space of Killing spinors is half that of
the vacuum solution, as anticipated.
The expression \onej\ for the Killing spinor $\epsilon$ can be rewritten as
$$
\epsilon = e^{{1\over2}\theta\gamma_9}H^{-{1\over8}}\epsilon_0\ ,
\eqn\conej
$$
where
$$
\tan\theta = H^{-{1\over2}}\tan\xi\ .
\eqn\abonej
$$
Note that these spinors vanish at the zeros of $H^{-1}$.

In order to show that the $SL(2;\R)$ transform of the solutions \dyons, for
which
$\langle \lambda\rangle \ne i$,  are also supersymmetric it suffices to show,
as pointed out previously in the context of D=4 dyons [\Ortin], that the
conditions
\Killing\ are $SL(2;\R)$ invariant. Let us denote by
$\hat {\cal D}(\lambda, {\cal F})$ the covariant derivative $\hat{\cal D}$ in
\oneh, thereby making explicit the dependence of this differential operator on
the
fields. Under the $SL(2;\R)$ transformation  of these fields,
$\lambda\rightarrow \lambda'$ and ${\cal F}\rightarrow {\cal F}'$ (given
explicitly in \onee\ and \onefb), one can show
that
$$
\hat {\cal D}(\lambda', {\cal F}') = e^{{1\over2}\phi\gamma_9} \hat {\cal
D}(\lambda, {\cal F}) e^{-{1\over2}\phi\gamma_9}
\eqn\inone
$$
where
$$
\tan\phi = {-ic(\lambda -\bar\lambda)\over 2d + c(\lambda +\bar\lambda) }\ ;
\eqn\intwo
$$
i.e. $\hat{\cal D}(\lambda, {\cal F})$ is an $Sl(2;\R)$-invariant covariant
derivatve.  If we take the $Sl(2;\R)$ transform of $\epsilon$ to be
$$
\epsilon' = e^{{1\over2}\phi\gamma_9}\epsilon\ ,
\eqn\infour
$$
then
$$
\hat {\cal D}(\lambda', {\cal F}')\epsilon' =
e^{{1\over2}\phi\gamma_9}\hat {\cal D}(\lambda, {\cal F})\epsilon\ ,
\eqn\ainfour
$$
Similarly, if $\chi(\lambda, {\cal F})$ is the spinor of \onei\ then
one can show that
$$
\chi(\lambda', {\cal F}') = e^{-{1\over2}\phi\gamma_9}\chi(\lambda, {\cal F})\
{}.
\eqn\inthree
$$
It follows that given background fields and a Killing spinor $\epsilon$
satisfying
the conditions \Killing\ for $\langle\lambda\rangle =i$, then the spinor
$$
\epsilon' = e^{{1\over2}(\theta+\phi)\gamma_9}H^{-{1\over8}}\epsilon_0
\eqn\ainthree
$$
satisfies the same conditions for the $Sl(2;\R)$ transformed solution with new
asymptotic value $\langle\lambda' \rangle \ne i$. Incidentally, this result
establishes the $Sl(2;\R)$ invariance of the modified Nester tensor $\hat
E^{\mu\nu}$  (assuming the above transformation property of
$\epsilon$) and the invariance of the Bogomolnyi bound is an immediate
consequence
of this.

%%%%%%%%%%%%%%%%%%%%%%%%%%%%%%%%%Chapter 4%%%%%%%%%%%%%%%%%%%%%%%%%%%%%%%%%%

\chapter{Singularity structure}

We now turn to the singularity structure of the dyonic membrane solutions
\dyons. Near a zero of $H^{-1}$ we have
$$
H \sim {\mu \over r^3}
\eqn\singone
$$
where
$$
r\equiv |{\bf x}-{\bf x}_n|\ .
\eqn\singtwo
$$
The asymptotic metric is
$$
r^{3\over2}(-dt^2+dy^2+dz^2) + {dr^2\over r^{3\over2}} + r^{1\over2}d\Omega_4^2
\eqn\singthree
$$
where $d\Omega_4^2$ is the metric on the unit 4-sphere.
One sees from this result that the proper distance to $r=0$ on a
surface of constant $t,y,z$ is finite, and that the radius of the four-sphere
of
constant $r$ on this surface shrinks to zero as $r\rightarrow 0$. It follows
that
the `lines' of force of $F$ must end on a singularity at $r=0$.

It is instructive to consider the membrane spacetime in the metric
$$
d\tilde s^2 = e^{2\sigma}ds^2\ ,
\eqn\abonej
$$
for which
$$
d\tilde s^2 = (\cos^2\xi + H^{-1}\sin^2\xi)[-dt^2 + dy^2 + dz^2] +
(\sin^2\xi + H\cos^2\xi)\; d{\bf x}\cdot d{\bf x}\ .
\eqn\athree
$$
The purely electric case now has a timelike naked singularity at zeros of
$H^{-1}$,
i.e. at a membrane core, so it would have to be identified with a
fundamental membrane. For this reason, one might choose to call the metric
$d\tilde s^2$ the `membrane metric'. Note that it would be the `string metric'
if
$\sigma$ were the dilaton, but $\sigma$ is {\it not} the dilaton.
In this `membrane metric' the metric for a membrane carrying magnetic
charge approaches the asymptotic metric
$$
d\tilde s^2 \sim \cos^2\xi\Big\{ [-dt^2 + dy^2 + dz^2] +
H\; d{\bf x}\cdot d{\bf x}\Big\}
\eqn\bthree
$$
near any of the membrane cores. Since $H\sim {\mu\over r^3}$ in this
limit, we now find that the proper distance to $r=0$ is infinite on a
hypersurface of constant $t,y,z$. Moreover, this remains true for timelike and
null
geodesics. Thus, the dyonic multi-membrane solutions are geodesically complete
{\it in the `membrane' metric} provided that the magnetic charge is non-zero.

Because $\sigma$ is not the dilaton, the interpretation of the above result
within
(type II) string theory is unclear. Moreover, since the dilaton has been set to
zero by the truncation, there is no longer any distiction between the Einstein
and
string metrics. Thus, the fact that our D=8 dyonic membrane solutions are
singular in
the Einstein metric implies that the string metric is also singular and this
must be
considered a difficulty in the context of type II superstring theory.
Fortunately,
this difficulty has a simple resolution if one considers the dyonic solutions
as
solutions of D=11 supergravity, which can be viewed as an effective action for
the
strongly coupled type IIA superstring [\PKTb,\Witten]. Consider the following
11-metric and four-form
$$
\eqalign{
ds_{11}^2 &= e^{{2\over3}\sigma} ds_8^2 + e^{-{4\over3}\sigma}\; d{\bf u}\cdot
d{\bf u}\cr
F_{11} &= F + 6du_1\wedge du_2\wedge du_3\wedge d\rho\ , }
\eqn\bfour
$$
where ${\bf u}$ are the coordinates of $T^3$ and $F$ is a field strength
four-form (F=dA) of the eight-dimensional spacetime. This field configuration
solves the equations of D=11 supergravity if the 8-metric, four-form
$F$, and scalar fields $\sigma$ and $\rho$ solve the D=8 field equations
\aonec.
This allows  us to lift the D=8 dyonic membrane solutions \dyons\ to D=11. The
result is
$$
\eqalign{
ds_{11}^2 &= H^{-{2\over3}}\Big[\sin^2\xi +H\cos^2\xi\Big]^{1\over3} (-dt^2 +
dy^2
+ dz^2) \cr
&+  H^{1\over3}\Big[\sin^2\xi +H\cos^2\xi\Big]^{1\over3} d{\bf x}\cdot
d{\bf x} + H^{1\over3}\Big[\sin^2\xi +H\cos^2\xi\Big]^{-{2\over3}} d{\bf
u}\cdot d{\bf u}\cr
F_{11} &=  {1\over2}\cos\xi (\star dH) +{1\over2}\sin\xi \; dH^{-1}\wedge
dt\wedge
dy\wedge dz\cr &\qquad -{3\sin 2\xi \over 2[\sin^2\xi +
H\cos^2\xi]^2}du_1\wedge
du_2 \wedge du_3\wedge dH\ .}
\eqn\bfive
$$

In the purely electric case, $\cos\xi=0$, we have
$$
\eqalign{
ds_{11}^2 &= H^{-{2\over3}}(-dt^2 + dy^2 + dz^2) + H^{1\over3}\big(d{\bf
x}\cdot
d{\bf x} + d{\bf u}\cdot d{\bf u}\big)\cr
F_{11} &= {1\over2}dH^{-1}\wedge dt\wedge dy\wedge dz\ . }
\eqn\bsix
$$
The harmonic function $H({\bf x})$ can now be interpreted as a harmonic
function
on $\E^5\times T^3$. The only difference between this solution of D=11
supergravity
and the multi-membrane solution found in [\DS] is that there $H$ was a harmonic
function on $\E^8$. Thus, the solution \bsix\ can be interpreted as a D=11
membrane in a background spacetime of topology $M_6\times T^3$ instead of
$M_{11}$, where $M_k$ indicates a $k$-dimensional Minkowski spacetime.

In the purely magnetic case, $\sin\xi=0$, we have
$$
\eqalign{
ds_{11}^2 &= H^{-{1\over3}}(-dt^2 + dy^2 + dz^2 + d{\bf u}\cdot d{\bf u} )
+ H^{2\over3}d{\bf x}\cdot d{\bf x}\cr
F_{11}&= {1\over2}\star dH\ , }
\eqn\bseven
$$
which is the fivebrane solution of D=11 supergravity [\Gu], except for the
periodic identification of the $T^3$ coordinates. We can therefore interpret
the
purely magnetic D=8 membrane as a D=11 fivebrane wrapped around a three-torus.
The D=11 multi-fivebrane solution of [\Gu] is geodesically complete [\GHT], the
singularities of $H$ being degenerate Killing horizons, so the singularity of
the
magnetic D=8 membrane solution is resolved by its interpretation in D=11, apart
from mild singularities introduced by the periodic identification of the $T^3$
coordinates.

These results for the purely electric and purely magnetic D=8 membranes confirm
the assumption made in [\PKTc] concerning their D=11 origin. Now we find that
the
more general dyonic membrane solution also has a D=11 interpretation. Although
the D=11 solution does not have an obvious $p$-brane interpretation, it is
non-singular, as we now show. Provided the magnetic charge is non-zero, i.e.
$\cos\xi\ne0$, the asymptotic form of the metric $ds_{11}^2$ of \bfive\ near
any
zero of $H^{-1}$ is
$$
ds_{11}^2 \sim (\cos\xi)^{2\over3}\Bigg\{ H^{-{1\over3}}(-dt^2 +dy^2 +dz^2 +
d{\bf
v}\cdot d{\bf v} ) + H^{2\over3}d{\bf x}\cdot d{\bf x}\Bigg\}\ ,
\eqn\beight
$$
where we have set ${\bf u}=(\cos\xi){\bf v}$. Apart from the overall factor the
result is independent of $\xi$. That is, the structure of the dyonic membrane
near
the singularities of $H$ is the same as for the purely magnetic case. We
conclude
that the singularities of the dyonic membranes are equally resolved in D=11.

%%%%%%%%%%%%%%%%%%%%%%%%%%%%%%%%%Chapter 5%%%%%%%%%%%%%%%%%%%%%%%%%%%%%%%%%%

\chapter{Comments}

In this paper we have obtained a bound on the tension of membrane solutions of
N=2 D=8 supergravity, and we have found the supersymmetric membrane solutions
that
saturate this bound. In general these solutions are dyonic. Since N=2 D=8
supergravity is obtained by a $T^3$ compactification of $D=11$ supergravity,
followed by a consistent truncation of the massive modes, the D=8 dyonic
membranes can be interpreted as solutions of D=11 supergravity. The purely
electric and purely magnetic D=8 membranes become the D=11 membrane and
fivebrane
respectively. The dyonic membranes have no obvious $p$-brane interpretation but
they are new solutions of D=11 supergravity which are non-singular if the
periodic identification of the $T^3$ coordinates (${\bf u}$) is relaxed. These
new solutions are intermediate between the D=11 membrane and fivebrane
solutions.
They might therefore be expected to play a role in the conjectured D=11
membrane/fivebrane duality [\HT,\PKTc].

The dyonic membrane solutions were given initially for a particular choice of
the
asymptotic values of the scalar fields that parameterise the possible vacua,
but they can then be found for any choice of vacuum by means of an $Sl(2;\R)$
transformation. In the context of type II string theory, an infinite set of
dyonic membrane solutions can be found, in equivalent vacua, by the action of
an
$Sl(2;\Z)$ subgroup of $Sl(2;\R)$ since this is a subgroup of the $SO(2,2;\Z)$
T-duality group. As explained in the introduction, this group can be
re-interpreted as the S-duality group of the equivalent heterotic string theory
after a compactification of the D=8 type II superstring to D=4 on $K_3$.
Some D=4 dyon solutions of the heterotic string will thereby acquire an
interpretation as D=8 dyonic membranes wrapped around the
homology two-cycles of $K_3$. These dyons {\it all} correspond to
non-perturbative
R-R states in the type II D=4 superstring but, according to the type II/
heterotic
equivalence conjecture, correspond to perturbative states of the heterotic
string
and their non-perturbative S-duals. In fact, they must include the dyons that
can
become massless at special points in the $K_3$ moduli space [\HTb], as expected
from the known symmetry restoration of the heterotic string at special points
in
its moduli space.

Dyonic membranes have many features in common with dyons. For example, let us
suppose that there is a purely magnetic membrane with charges $(p,q)=(1,0)$
when $\langle\lambda\rangle=i$; this amounts to the assumption that, in this
vacuum, the choice of
$\xi=0$ in \dyons\ is admissable in the quantum theory. Now consider a new
vacuum
related to the original one by an
$Sl(2;\R)$ transformation with the element
$$
\pmatrix{a&b\cr c&d} = \pmatrix{1&b\cr0&1}\ .
\eqn\dyonsfive
$$
One finds that $\langle\lambda'\rangle = b+ i$, or equivalently
$\langle\sigma\rangle =0$, $2\langle\rho\rangle=b$, in the new vacuum and
that the membrane solution in this vacuum has charges $(p,q)= (1,b)=
(1,2\langle\rho\rangle)$. Thus, a dyonic membrane with unit magnetic charge has
a
fractional electric charge given by
$$
q= 2\langle\rho\rangle
\eqn\newone
$$
This is just the generalization to dyonic membranes of the Witten effect for
dyons
[\Witb]. The identification of vacua related by an $Sl(2;\Z)$ tansformation
implies, in particular, that $2\rho\cong 2\rho +1$, so the value of $q$ for a
dyon with unit magnetic charge will change by one as the asymptotic value of
$2\rho$ is smoothly continued from $2\langle\rho\rangle$ to
$2\langle\rho\rangle
+1$. In the D=4 dyon case, this continuation of $\langle\rho\rangle$ can be
realized physically by transport around an axion string. In the D=8 dyonic
membrane case it could be achieved by transport around an axionic fivebrane.

There is, however, a new feature of dyonic membranes not shared by dyons. To
see
this, we note that given the existence of a particle with charges $(0,1)$ in
the
vacuum with $\lambda=i$, the DSZ quantization condition implies that for any
other
particle with charges $(p,q)$, necessarily $p\in \Z$, i.e. while electric
charge
can be fractional, magnetic charge cannot be. Had we assumed the existence of a
particle with charges $(1,0)$ we would have instead deduced that $q\in \Z$ and
$p$
could be fractional. The DSZ quantization condition does not distinguish
between
these possibilities, but perturbation theory does: in string perturbation
theory
there exist particles with only electric charge and all semi-classical dyons
have
integer magnetic charge. A similar conclusion can be made for any of the vacua
in
the same equivalence class of $\lambda=i$; as we saw earlier for dyonic
membranes, the assumption that there exist purely electric solutions is
equivalent to the assumption that $\tan\xi$ is rational. It seems, therefore,
that
for dyons the appeal to perturbation theory allows us to restrict the allowed
values of the angular parameter analogous to $\xi$, but the same does
not apply to dyonic membranes, at least in the context of type II superstring
theory, because all membrane solutions, electric, magnetic or dyonic, are
non-perturbative.


\vskip 1cm
\noindent{\bf Acknowledgements:} G.P. was supported by a Royal Society
University
Research Fellowship. J.M.I thanks the Commission of the European Community and
CICYT (Spain) for financial support. We thank C.M. Hull for helpful
discussions.

\refout

\bye